\documentclass[aps,prl,showpacs,twocolumn,groupedaddress]{revtex4}

\usepackage{graphicx}  % needed for figures
\usepackage{dcolumn}   % needed for some tables
\usepackage{bm}        % for math
\usepackage{amssymb}   % for math

\def\lsim{\mathrel{\rlap{\lower4pt\hbox{\hskip1pt$\sim$}}\raise1pt\hbox{$<$}}}
\def\gsim{\mathrel{\rlap{\lower4pt\hbox{\hskip1pt$\sim$}}\raise1pt\hbox{$>$}}}

\begin{document}

\hspace{5.2in} \mbox{FERMILAB-PUB-08-154-E}

\title{Search for a scalar or vector particle decaying into $Z\gamma$ in $p\bar{p}$ collisions at $\sqrt{s}$~=~1.96~TeV}

% LIST_OF_AUTHORS_R2.TEX               5/23/08              
%
\author{V.M.~Abazov$^{36}$}
\author{B.~Abbott$^{75}$}
\author{M.~Abolins$^{65}$}
\author{B.S.~Acharya$^{29}$}
\author{M.~Adams$^{51}$}
\author{T.~Adams$^{49}$}
\author{E.~Aguilo$^{6}$}
\author{M.~Ahsan$^{59}$}
\author{G.D.~Alexeev$^{36}$}
\author{G.~Alkhazov$^{40}$}
% ...

\affiliation{\vspace{0.1 in}(The D\O\ Collaboration)\vspace{0.1 in}}
% ...
\affiliation{$^{6}$University of Alberta, Edmonton, Alberta, Canada,
                Simon Fraser University, Burnaby, British Columbia, Canada,
                York University, Toronto, Ontario, Canada, and
                McGill University, Montreal, Quebec, Canada}
% ...
\affiliation{$^{29}$Tata Institute of Fundamental Research, Mumbai, India}
% ...
\affiliation{$^{36}$Joint Institute for Nuclear Research, Dubna, Russia}
% ...
\affiliation{$^{40}$Petersburg Nuclear Physics Institute,
                St. Petersburg, Russia}
% ...
\affiliation{$^{49}$Florida State University, Tallahassee, Florida 32306, USA}
% ...
\affiliation{$^{51}$University of Illinois at Chicago,
                Chicago, Illinois 60607, USA}
% ...
\affiliation{$^{59}$Kansas State University, Manhattan, Kansas 66506, USA}
% ...
\affiliation{$^{65}$Michigan State University,
                East Lansing, Michigan 48824, USA}
% ...
\affiliation{$^{75}$University of Oklahoma, Norman, Oklahoma 73019, USA}
% ...
  % input Dzero author list

% Use floating date so far, when publishing - freeze the date
\date{June 3, 2008}

\begin{abstract}
We present a search for a narrow scalar or vector resonance decaying into $Z\gamma$ with a
subsequent $Z$ decay into a pair of electrons or muons. The data for this search were collected with the D0 detector at the Fermilab
Tevatron $p\bar{p}$ collider at a center of mass energy $\sqrt{s}$ = 1.96 TeV.
Using 1.1 (1.0) fb$^{-1}$ of data, we observe 49 (50) candidate events in the electron (muon) channel, in good agreement with the standard model prediction. From the combination of both channels, we derive 95\% C.L. 
upper limits on the cross section times branching fraction ($\sigma \times {\cal B}$) into $Z\gamma$. 
These limits range from 0.19 (0.20) pb for a scalar (vector) resonance mass of 600 GeV/$c^2$ to 2.5 (3.1) pb for a 
mass of 140 GeV/$c^2$.
\end{abstract}

\maketitle

%%%%%%%%%%%%%%%%%%%%%%%%%%%%%%%%%%%%%%%%%%%%%%%%%%%%%%%%%%%%%%%%%%%%%%%%%%%
Despite its tremendous success, the standard model (SM) in its current form may be a low energy approximation of a more fundamental theory. The SM does not describe gravity, and fundamental parameters such as masses and coupling constants are not derived from the theory. Many models exist to replace or extend the SM. A heavy partner of the $Z$ boson, $Z^{\prime}$, appears in grand unified theories, little Higgs models, models with extra spatial dimensions, and superstring theories. Scalar Higgs bosons, pseudo-scalar toponium, vector $Z^{\prime}$ bosons, and techniparticles could decay into the diboson final state $Z \gamma$~\cite{Buescher:2005re,Djouadi,Kozlov:2005rj,Ono:1983tf,Cakir:2004nh,Hill}.

This Letter presents a search for a narrow scalar or vector resonance decaying into $Z\gamma$ using approximately 1 fb$^{-1}$ of data collected with the D0 detector in $p\bar{p}$ collisions at $\sqrt{s}$ = 1.96 TeV at the Fermilab Tevatron collider. This analysis considers leptonic decays of the $Z$ boson into electron or muon pairs. A similar search had been carried out by the D0 collaboration using a smaller dataset corresponding to about 300 pb$^{-1}$~\cite{us,Abazov:2006ez}. 

\begin{figure}
\begin{center}
$$
\begin{array}{cc}
\includegraphics[scale=0.2]{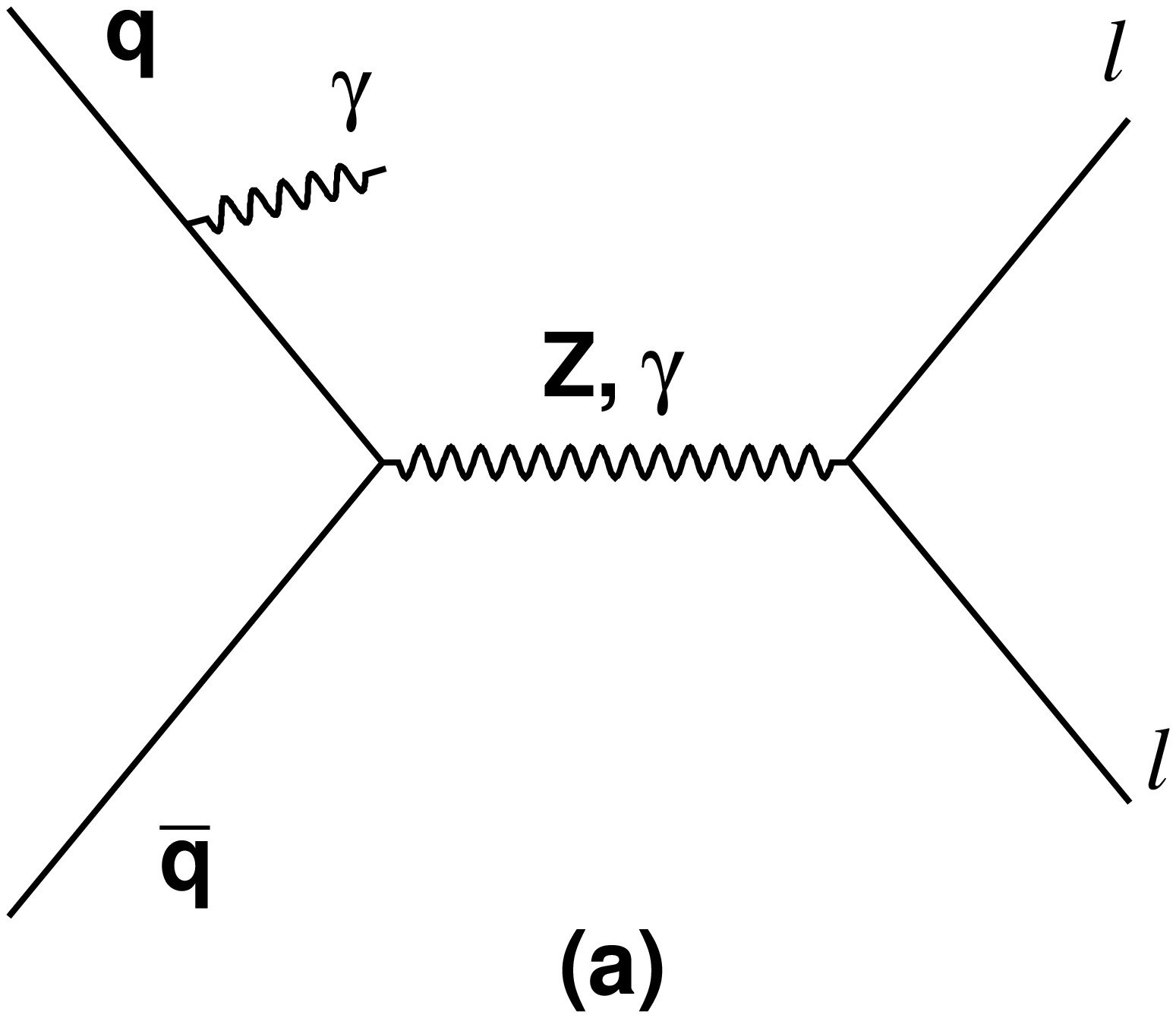} &
\includegraphics[scale=0.2]{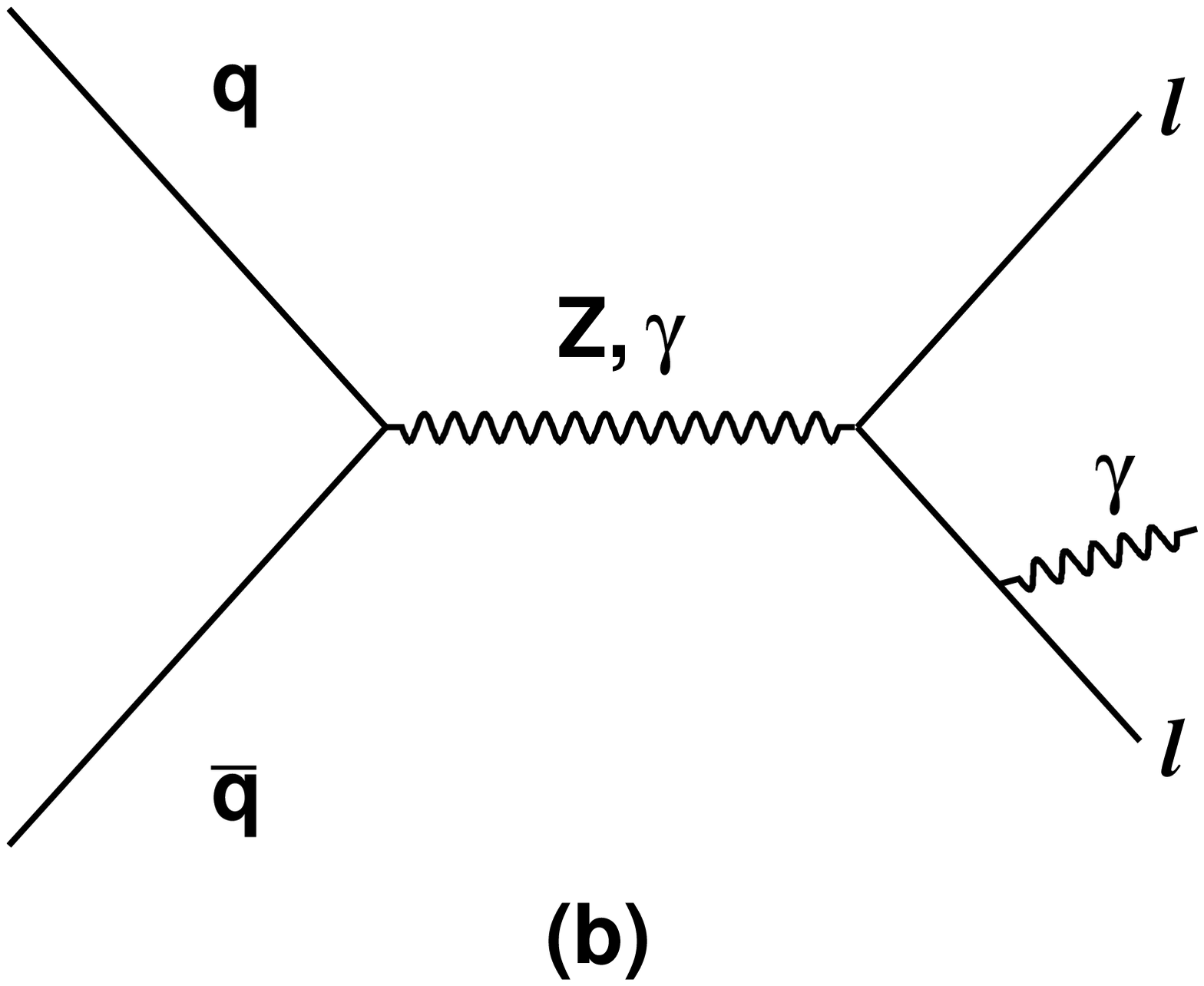} \\
\includegraphics[scale=0.2]{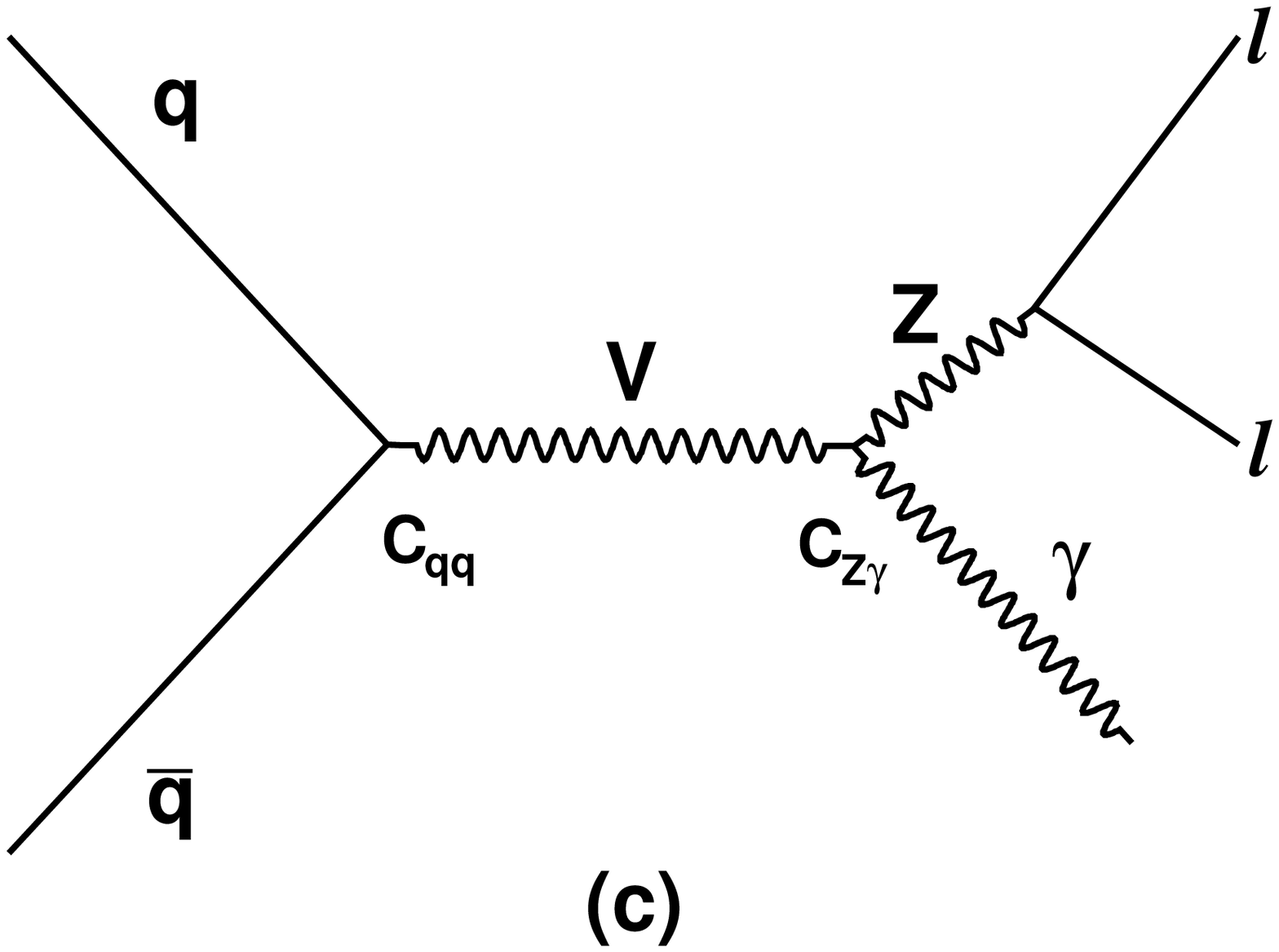} &
\includegraphics[scale=0.2]{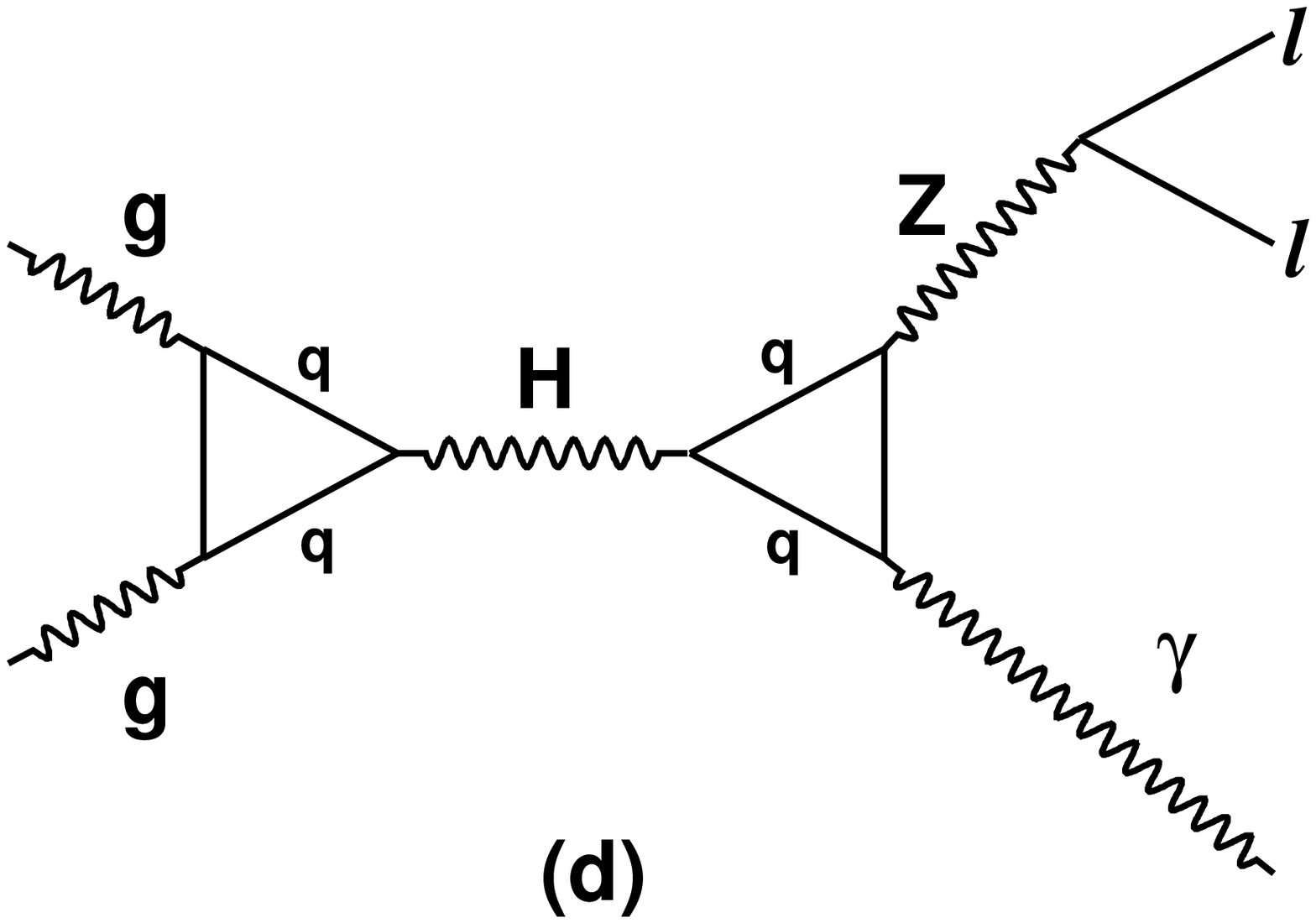}
\end{array}
$$
\caption{Diagrams for the leading-order processes which produce $Z\gamma$ candidates:
(a) SM initial state radiation (ISR), (b) SM final state radiation (FSR), (c)
$q \bar q$ pair annihilation into a vector ($V$) particle which couples to the $Z \gamma$ and (d) SM Higgs production and decay.
\label{fig:feyn}}
\end{center}
\end{figure}

The main components of the D0 detector \cite{run2det} include a central tracking system, a sampling calorimeter, and a muon detection system. The central tracker consists of a silicon microstrip tracker (SMT) and a central fiber tracker (CFT), both surrounded by a 2 T superconducting solenoid. These two subsystems provide tracking and vertexing for charged particles with pseudorapidities up to $|\eta| \approx$~3 and $|\eta| \approx$~2.5, respectively. The pseudorapidity is defined as $\eta = -\ln[\tan (\theta/2)]$, where $\theta$ is the polar angle in the D0 coordinate system with the origin at the geometrical center of the detector, the $z$-axis pointing along the beam line and the $y$-axis vertical upward. The liquid-argon/uranium
calorimeter is spatially divided into the central calorimeter (CC, $|\eta| < $~1.1) and two end calorimeters 
(EC, 1.5~$ < |\eta| < $~4.2). Both the CC and EC are segmented into electromagnetic (EM) and hadronic (HAD)
sections. The outermost subdetector, the muon system, consists of three layers of tracking detectors and scintillation
trigger counters, and a 1.8~T toroidal iron magnet between the first two inner layers. The muon system is capable
of providing measurements for particles with pseudorapidities $|\eta| <$~2. Arrays of plastic scintillators situated near the beamline in
front of the EC are used to measure the instantaneous luminosity of the colliding proton--antiproton beams. Events in the electron (muon) channel are collected using a suite of single electron (single or dimuon) triggers, corresponding to an integrated luminosity of 1.1 (1.0) fb$^{-1}$. The uncertainty on the luminosity is 6.1\% \cite{lumi}. 

A scalar model and a vector model for a particle decay into $Z\gamma$ are used; both assume a resonance total width
 smaller than the resolution of the detector (Table~\ref{tab:muon_constraint}). In this narrow-width approximation, interference effects are negligible \cite{Carena}. We use the SM Higgs boson production model, as 
implemented in {\sc pythia}~\cite{pythia}, for the scalar resonance decay into the $Z\gamma$ final state. To model the  vector resonance decay, we use a generic color-singlet, neutral, vector particle ($V$) implemented in {\sc madevent} \cite{madevent}. We assume $C_{qq}$ is the coupling between the $V$ and initial state fermions $q \overline{q}$, with $q=u$ or $d$ and that $C_{Z\gamma}$ is the coupling between the $V$ and $Z \gamma$, as shown in Fig.~\ref{fig:feyn}. A
$Z^{\prime}$ is a good example of a $V$ particle, but there is no
model of fundamental $Z^{\prime}$ coupling to $Z\gamma$, since the $Z^{\prime}$ has no
electric charge. However, if the $Z^{\prime}$ has a composite structure, as in technicolor models, then such a decay is possible.   

Electromagnetic objects such as electrons and photons are required to have an isolation~\cite{EM_isolation} of less than 0.2 and 0.15, respectively. Electron candidates from the $Z$ boson decay are reconstructed from EM showers in the calorimeter with an electron-like shower shape, that are required to satisfy the following criteria: have transverse
energy $E_T >$~15~GeV, deposit at least 90\% of their energy in the EM calorimeter, and have tracks spatially matched to the EM showers. At least one electron candidate is required to be
reconstructed in the central calorimeter, while the second candidate can be reconstructed either in the CC or EC. In addition, at least one electron must have $E_T >$~25~GeV to satisfy single-electron high-$E_T$ triggers.

Reconstruction of the $Z \to \mu\mu$ decays begins with a search for a pair of muons with transverse momentum $p_T >$~15~GeV/$c$. To reduce the effects of the muon trigger $p_T$ turn-on, at least one muon must have $p_T >$~20~GeV/$c$. Cosmic-ray background is reduced by rejecting muon candidates that do not originate from the same vertex or are reconstructed back-to-back with an opening angle $|\Delta\phi_{\mu\mu} + \Delta\theta_{\mu\mu} - 2\pi|$ less than 0.05, where $\Delta\phi_{\mu\mu}$ and $\Delta\theta_{\mu\mu}$ are the muon candidates separations in polar and azimuthal angles. The azimuthal angle is defined as $\phi = \arctan (\frac{y}{x})$ so that $\phi=\frac{\pi}{2}$ points along the $y$-axis. Contamination from hadronic $b \bar b$ production is
reduced by the additional requirement that one or both of the muon candidates are isolated from other activity in the calorimeter and central tracker. For calorimeter isolation the $E_T$ sum of calorimeter cells within an $(\eta,\phi)$ annulus centered on the muon trajectory with $R = \sqrt{(\Delta \phi)^{2} + (\Delta \eta)^{2}}$ between 0.1 and 0.4 should be less than 2.5 GeV. For central tracker isolation the sum of the transverse momenta of all tracks reconstructed within a cone centered on the muon trajectory with $R$ = 0.5 should be less than 3.5 GeV/$c$. The transverse momentum 
of the muon track itself is not included in this sum.

The $Z\gamma$ final state is obtained by requiring an event to also have a photon candidate, reconstructed from an isolated
shower in the EM part of the central calorimeter with the shower shape consistent with that of a photon. The photon candidate must be separated from both leptons from the $Z$ boson decay by $\Delta R >$~0.7. A jet with most of its energy carried by photons, and
mis-identified as an electron or a photon candidate is defined as an EM-like jet. EM-like jet background is reduced by requiring that the sum of transverse
momenta of all tracks, reconstructed within an $(\eta,\phi)$ annulus centered in a photon candidate's trajectory, is below 1.5 GeV/$c$. The annulus has a radius $R$ between 0.05 and 0.4.

The three-body mass ($M_{\ell\ell\gamma}$) resolution directly affects the sensitivity in searching for a narrow mass resonance. The $M_{\ell\ell\gamma}$ resolution is 8--18\% in the muon channel and 4--5\% in the electron channel for SM sources. To improve the resolution for the reconstructed three-body invariant mass in the muon channel, a $Z$ mass constraint is applied  to adjust the muon transverse momenta
in the dimuon channel. A $\chi^2$ function is minimized by varying the muons' $p_T$ with the dimuon mass constrained to the $Z$ mass; the fit has one degree of freedom. If the $\chi^2$ after minimization, $\chi^2_{\rm min}$, is less than seven, the constrained fit is used for the $Z$ momentum and mass. In cases where the constraint $\chi^2_{\rm min}$ is greater than seven, the mass constraint is not applied. The choice of $\chi^2_{\rm min} <$ 7 is found by scanning values of $\chi^2$ and calculating the acceptance for the $\mu\mu\gamma$ final state. This requirement ensures that Drell-Yan events will only be moved to the $Z$ boson mass
if they are consistent with that mass. In Monte Carlo studies this technique improves the $M_{\ell\ell\gamma}$ resolution substantially in the muon channel, as shown in Table~\ref{tab:muon_constraint}.
\begin{table}[ht!]
\begin{center}
\caption{Dilepton-plus-photon invariant mass resolution. The fourth column shows an improvement in the dimuon-plus-photon invariant mass resolution after the $Z$ mass constraint has been applied.
\label{tab:muon_constraint}}
\begin{tabular}{cccc}
\hline
\hline
 &  $M_{ee\gamma}$  & $M_{\mu\mu\gamma}$ & $M_{\mu\mu\gamma}$   \\
Mass (GeV/$c^2$)  & resolution & uncorrected  & corrected \\
  & ($\%$) & resolution ($\%$) & resolution ($\%$)\\
\hline
100     & 4.1  & 8.3    &   4.6  \\
200     & 4.4  & 8.0    &   4.7  \\
300     & 4.8  & 9.9    &   5.2  \\
400     & 4.9  & 11.7   &   5.5  \\
500     & 4.9  & 11.8   &   5.8  \\
600     & 4.9  & 13.6   &   6.1  \\
700     & 4.9  & 14.6   &   6.4  \\
800     & 5.2  & 16.8   &   7.2  \\
900     & 5.3  & 17.6   &   7.8  \\
\hline
\hline
\end{tabular}
\end{center}
\end{table}

To improve the analysis sensitivity in an unbiased fashion, an optimization of the photon 
$E_T$~and dilepton invariant mass $M_{\ell\ell}$ selection criteria
is performed with respect to $S/\sqrt{S+B}$ using a simulation. Here $S$ is the number of signal events and $B$ is 
the number of background events. The simulated signal is based on MC samples of vector and scalar resonances decaying to 
the $Z\gamma$ final state. The two dominant background sources, SM $Z\gamma$ and $Z$+jets production, are considered. 
The background normalization procedure is discussed in detail below. 
The results for optimization varied with the channel and resonance mass.
Because the $S/\sqrt{S+B}$ has a broad maximum and the
$M_{\ell\ell\gamma}$ distribution still had substantial discrimination, selection criteria are chosen that allowed greater acceptance. The final conditions imposed are $E_{T}^{\gamma}~>~$20~GeV and dilepton mass $M_{\ell\ell}~>~$80~GeV/$c^2$.

The photon and lepton reconstruction efficiencies, as well as
the total acceptances for different $Z$ boson
decay modes (electron and muon), depend on the spin and the mass of
the hypothetical resonance. Electron and muon decay modes are also treated separately to take into
account differences in geometrical acceptance, trigger and reconstruction efficiencies of electrons and muons.
The efficiency to reconstruct a pair of electrons for resonances with invariant masses from 120--900 GeV/$c^2$ 
varies between 60--68\% (61--67\%) for a vector (scalar) resonance. The photon reconstruction
efficiency varies from 92\% to 95\% in both electron and muon channels for both types
of resonances. The electron channel efficiencies rise slowly by about a factor of two up to 
resonance masses of 600~GeV/$c^2$, whereupon they begin to drop. Similar effects are observed in the 
muon channel. The muon identification efficiency is approximately 79\% per a pair of muons for all resonance masses in either model. Single electron triggers are $(99 \pm 1)\%$ efficient,
and the average efficiency for the muon trigger requirements is $(68 \pm 1)\%$. Both the total efficiency of the event selection criteria multiplied by the geometrical and kinematic 
acceptance, and the trigger efficiency have a noticeable mass dependence, rising from 7\% to 19\%  for vector resonance masses between 120~GeV/$c^2$ and 600~GeV/$c^2$,
and from 8\% to 20\% for scalar resonance masses over the same interval in the electron channel. Similar effects are observed in the muon channel. A source of inefficiency appears at masses above 600~GeV/$c^2$ where the leptons 
are spatially more collinear and can become indistinguishable. For this reason we require the di-electron pair to be separated by $\Delta R_{ee} > 0.6$, while muon separation
$\Delta R_{\mu \mu}$ is above 0.5.

The two main background sources to the process under study are SM $Z\gamma$
production and $Z$+jet production, where a jet is misidentified as a photon. All other background sources were found to be negligible. To estimate the  $Z$+jet background, we first calculate the $E_T$-dependent
rate, $f$, at which an EM-like jet is misreconstructed as a photon. This is done using a sample of events enriched with jets that satisfy the jet trigger requirements.
The rate is the ratio of the $E_T$ spectrum of photon candidates that pass all
photon selection criteria and the $E_T$ spectrum of EM objects that are reconstructed in the geometrical acceptance of the central calorimeter. The rate is further corrected for contamination from direct photon production. To estimate the $Z$+jet background two samples are used. One sample is our final data sample that contains events with a $Z$ boson candidate and a photon candidate that passes all selection criteria and a sample, referred as to $Z$+EM, consisting of events that contain a $Z$ boson candidate plus a photon candidate that fails the track isolation and shower shape requirements. The latter sample comprises of real photons and EM-like jets. Our final data sample contains $Z+\gamma$ events that are corrected by the photon identification efficiency $\epsilon_\gamma$ and $Z$+jet events that are corrected by $f$. More details on the calculation of the $Z$+jet background can be
found in Ref.~\cite{us}.
The final $Z$+jet background is estimated to be $4.5 \pm 0.7({\rm stat.}) \pm 0.6({\rm syst.})$ events in the electron channel and $4.4 \pm 0.7({\rm stat.}) \pm 0.6({\rm syst.})$ events in the muon channel. The systematic uncertainty on the $Z$+jet background mostly comes from the uncertainty on the photon efficiency and the rate at which an EM-like jet is misreconstructed as a photon.

The number of SM $Z\gamma$ background events is estimated from a $Z\gamma$ MC sample obtained with the leading-order (LO) Baur event generator~\cite{baur} using CTEQ6L1 parton distribution functions (PDFs) with values of zero for the trilinear $Z\gamma\gamma$ and $ZZ\gamma$ couplings: $N_{\rm SM}(Z\gamma \rightarrow \ell\ell\gamma) = \epsilon_{tot}\cdot \sigma_{\rm SM}(Z\gamma)\cdot {\cal B}(Z \rightarrow \ell\ell)\cdot {\cal L}$. Here, $\epsilon_{tot}$ is the total efficiency, $\sigma_{\rm SM}(Z\gamma)\cdot {\cal B}(Z \rightarrow \ell\ell)$ is the
cross section times branching fraction and ${\cal L}$ is the integrated luminosity. We correct the LO photon $E_T$~spectrum using the $E_T$-dependent $K$-factor derived from the next-to-leading-order (NLO) Baur event generator~\cite{baur_NLO}. Table~\ref{tab:SMZg_bkg} summarizes these numbers~\cite{us}. Using the above equation we obtain an estimated SM $Z\gamma$ contribution of $37.4 \pm 6.1({\rm stat.}) \pm 2.6({\rm syst.})$ events in the electron channel and $41.6 \pm 6.5({\rm stat.}) \pm 2.2({\rm syst.})$ events in the muon channel. The systematic uncertainty on the SM $Z\gamma$ background comes from the 
uncertainty on the theoretical cross section, the PDFs and reconstruction efficiency.

\begin{table}[ht!]
\begin{center}
\caption{Summary of the components used to estimate number of SM $Z\gamma$ background events.
\label{tab:SMZg_bkg}}
\begin{tabular}{ccc}
\hline
\hline
Parameter &  Electron channel  & Muon channel\\
\hline
$\epsilon_{tot}$        & $0.0026 \pm 0.0001$  & $0.0032 \pm 0.00007$  \\
$\sigma_{\rm SM}(Z\gamma)\cdot {\cal B}(Z \rightarrow \ell\ell)$ (pb) & $12.85 \pm 0.60$     & $12.85 \pm 0.60$  \\
${\cal L}$ (pb$^{-1}$)                    & $1110 \pm 70$        & $1010 \pm 60$ \\
\hline
\hline
\end{tabular}
\end{center}
\end{table}

The selection criteria yield 49 candidates in the electron channel and 50 candidates in the muon channel 
with the estimated combined SM $Z\gamma$ plus $Z$+jet background of $41.9 \pm 6.2({\rm stat.}) \pm 2.6({\rm
 syst.})$ events in the electron channel and $46.0 \pm 6.6({\rm stat.}) \pm 2.3({\rm syst.})$ events
in the muon channel, as shown in Table~\ref{tab:summary}. 

\begin{table}[ht!]
\begin{center}
\caption{Summary of the background expectations in each channel 
  and comparison with the observation. The first uncertainty is statistical and the second is systematic.
\label{tab:summary}}
\begin{tabular}{ccc}
\hline
\hline
    &  Electron channel  & Muon channel\\
\hline
SM $Z\gamma$ & $37.4 \pm 6.1 \pm 2.6$  &  $41.6 \pm 6.5 \pm 2.2$  \\
Z + jets  &  $4.5 \pm 0.7 \pm 0.6$    & $4.4 \pm 0.7 \pm 0.6$ \\
Total background        & $41.9 \pm 6.2 \pm 2.6$  & $46.0 \pm 6.6 \pm 2.3$  \\
Data        & 49  &  50 \\
\hline
\hline
\end{tabular}
\end{center}
\end{table}

\begin{figure}
\begin{center}
\includegraphics[scale=0.36]{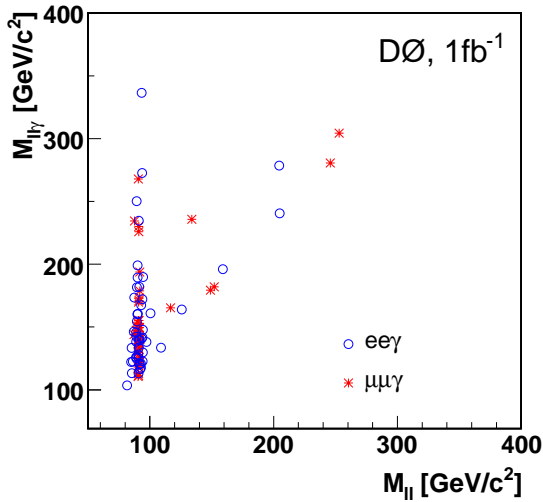}
\caption{Invariant mass of the dilepton system vs. invariant mass of
dilepton-plus-photon candidates.
\label{fig:mll_mllg}}
\end{center}
\end{figure}

Figure~\ref{fig:mll_mllg} shows the distribution of the dilepton invariant mass ($M_{\ell\ell}$) versus the dilepton-plus-photon invariant mass ($M_{\ell\ell\gamma}$). The vertical band is populated by the ISR events where the radiated photon 
originates from one of the initial partons and the on-shell $Z$ boson 
decays into two leptons. The Drell-Yan events cluster along the diagonal band.
Most of the FSR events, which would populate the horizontal band centered at 
$M_{\ell\ell\gamma} \approx M_Z$, are removed by the $M_{\ell\ell} >$~80 GeV/$c^2$ cut.

\begin{figure}
\begin{center}
\includegraphics[scale=0.36]{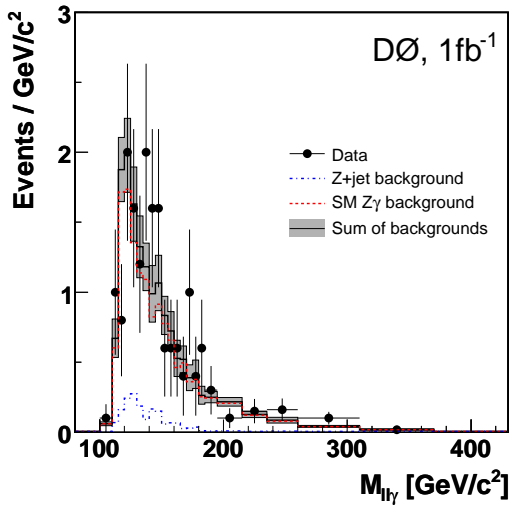}
\caption{
Invariant dilepton-plus-photon mass spectrum for $\ell\ell\gamma$ data (dots), 
SM $Z\gamma$ background (solid line histogram) and $Z$+jet background (dashed line histogram). The shaded band illustrates the systematic and statistical uncertainty on the sum of backgrounds.
\label{fig:mllg_vector}}
\end{center}
\end{figure}

\begin{figure}
\begin{center}
\includegraphics[scale=0.36]{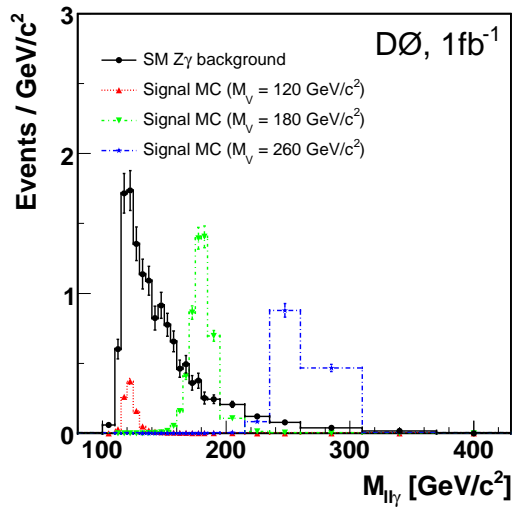}
\caption{Shape comparison of the invariant dilepton-plus-photon mass spectrum associated with MC signal of a vector particle decaying into $Z\gamma$ for vector resonance masses of 120, 180 and 260~GeV/$c^2$.
\label{fig:mllg_vector_various}}
\end{center}
\end{figure}

In Fig.~\ref{fig:mllg_vector}, the combined $M_{\ell\ell\gamma}$ distribution from both channels is compared with the SM background. Due to the limited available background statistics and the three-body mass resolution, events with $M_{\ell\ell\gamma} >$~370~GeV/$c^2$ are placed into an overflow bin. Figure~\ref{fig:mllg_vector_various} shows the $M_{\ell\ell\gamma}$ distribution associated with MC signals of a vector particle decaying into $Z\gamma$ for different vector resonance masses. 

The observed $M_{\ell\ell\gamma}$ spectrum is found to be consistent with SM
expectations, hence limits are set on the $\sigma \times {\cal B}$ for both vector and scalar models. The branching fraction for $Z$ to $ee$ or $\mu \mu$ is accounted for in these results. 

\begin{figure}
\begin{center}
\includegraphics[scale=0.45]{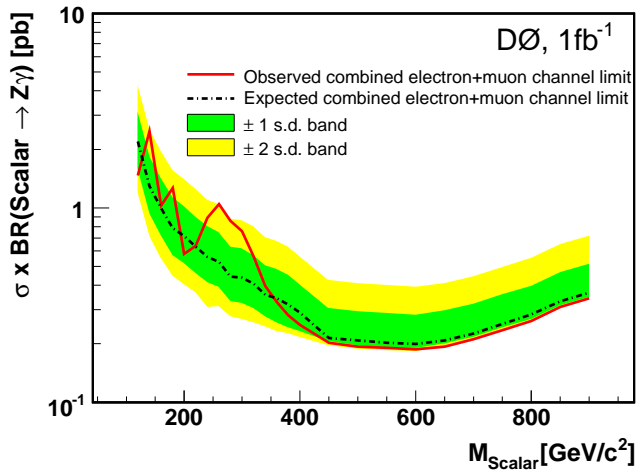}
\caption{The observed $\sigma \times {\cal B}$ 95\% confidence level limit for a scalar
particle decaying into Z$\gamma$ as a function of the scalar resonance mass. The observed limit is compared to the expected limit for a SM Higgs decaying into Z$\gamma$. The two shaded bands represents the 1 s.d. (dark) and 2 s.d. (light) uncertainties on the expected limit.
\label{fig:scalar_limit}}\end{center}\end{figure}

\begin{figure}
\begin{center}
\includegraphics[scale=0.45]{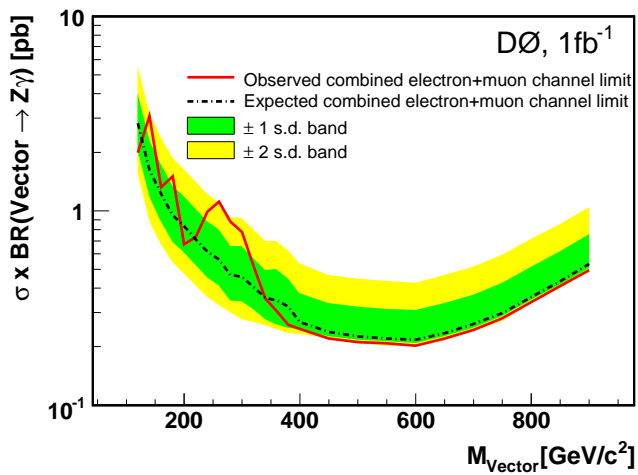}
\caption{The observed $\sigma \times {\cal B}$ 95\% confidence level limit for a vector
particle decaying into Z$\gamma$ as a function of the vector resonance mass. The observed limit is compared to the expected limit for a generic color-singlet, charge-singlet, vector particle decaying into Z$\gamma$. The two shaded bands represents the 1 s.d. (dark) and 2 s.d. (light) uncertainties on the expected limit.
\label{fig:vector_limit}}
\end{center}
\end{figure}

A modified frequentist method~\cite{fisher} is used to examine the $M_{\ell\ell\gamma}$ spectrum in the data (Fig.~\ref{fig:mllg_vector}) for 
discrepancies with respect to SM sources. A Poisson log-likelihood ratio test statistic (LLR)~\cite{junk} is used 
to compare the SM-only background hypothesis to one that incorporates a possible $Z\gamma$ resonance signal. 
The LLR incorporates systematic uncertainties in the form of nuisance parameters that are integrated out assuming 
a Gaussian prior and a relative contribution to the signal and background uncertainties that is independent of the 
$M_{\ell\ell\gamma}$ invariant mass. When setting the limits using the LLR method in the combined electron and muon channels, a 2.3\% reconstruction efficiency times acceptance systematic uncertainty is applied to the MC signal. 
A 6.1\% systematic uncertainty from luminosity and 5\% PDF uncertainty are applied to the signals and SM $Z\gamma$ background. An additional systematic uncertainty of 9\% on the $Z$+jet background is due to the photon efficiency and the rate at which an EM-like jet is misreconstructed as a photon, whereas an additional systematic uncertainty of 2.6\% on the SM $Z\gamma$ background is due to the theoretical cross section and reconstruction efficiency times acceptance.
Figures~\ref{fig:scalar_limit} and \ref{fig:vector_limit} show 95\% C.L. exclusion curves for $\sigma \times {\cal B}$ as function of the resonance mass in the vector and scalar models, respectively.


In summary, we have searched for evidence of a narrow $Z\gamma$ resonance in $ee\gamma$ and $\mu\mu\gamma$ final states of $p \bar p$ collisions at $\sqrt{s} = 1.96$ TeV using data collected by the D0 detector during the 2002--2006
run of the Fermilab Tevatron $p\bar{p}$ Collider. We observe 49 candidate events in the electron channel and 50 in the muon channel,
consistent with expectations from SM processes.
No statistically significant evidence of a resonance decaying into $Z\gamma$ is
observed. From the combination of both channels, we derive 95\% C.L. upper limit on cross section times branching fraction, which ranges from 0.19 (0.20) pb for a scalar (vector) resonance mass of 600 GeV/$c^2$ to 2.5 (3.1) pb for a mass of 140 GeV/$c^2$.

%%%%%%%%%%%%%%%%%%%%%%%%%%%%%%%%%%%%%%%%%%%%%%%%%%%%%%%%%%%%%%%%%%%%%%%%%%%%
We would like to thank Steve Mrenna for providing us with the adapted {\sc Madevent} generator. 
% acknowledgement_paragraph_r2.tex                         5/23/08
%
We thank the staffs at Fermilab and collaborating institutions, 
and acknowledge support from the 
DOE and NSF (USA);
CEA and CNRS/IN2P3 (France);
FASI, Rosatom and RFBR (Russia);
CNPq, FAPERJ, FAPESP and FUNDUNESP (Brazil);
DAE and DST (India);
Colciencias (Colombia);
CONACyT (Mexico);
KRF and KOSEF (Korea);
CONICET and UBACyT (Argentina);
FOM (The Netherlands);
STFC (United Kingdom);
MSMT and GACR (Czech Republic);
CRC Program, CFI, NSERC and WestGrid Project (Canada);
BMBF and DFG (Germany);
SFI (Ireland);
The Swedish Research Council (Sweden);
CAS and CNSF (China);
and the
Alexander von Humboldt Foundation (Germany).


\begin{thebibliography}{99}
% list_of_visitor_addresses_r2.tex                         5/23/08
%  available symbols are:
%  $\ast, \dag, \ddag, \S, \P, $\|$, $\ast\ast$, \dag\dag, \ddag\ddag ,\#
%
\bibitem[a]{alton}
Visitor from Augustana College, Sioux Falls, SD, USA.
\bibitem[b]{burdin}
Visitor from The University of Liverpool, Liverpool, UK.
\bibitem[c]{podesta-lerma}
Visitor from ICN-UNAM, Mexico City, Mexico.
\bibitem[d]{quadt,meyer,hensel,park}
Visitor from II. Physikalisches Institut, Georg-August-University, G{\"o}ttingen, Germany.
\bibitem[e]{voutilainen}
Visitor from Helsinki Institute of Physics, Helsinki, Finland.
\bibitem[f]{wenger}
Visitor from Universit{\"a}t Z{\"u}rich, Z{\"u}rich, Switzerland.
%\bibitem[?]{coadou}
%Visitor from Simon Fraser University, Vancouver, B.C., Canada.
%\bibitem[?]{kozminski}
%Visitor from Lewis University, Romeoville, IL, USA.
\bibitem[\ddag]{deceased}
Deceased.

%
\vskip 0.25cm
  
\bibitem{Buescher:2005re} V.~Buescher and K.~Jakobs, Int.\ J.\ Mod.\ Phys.\ A {\bf 20}, 2523 (2005).
\bibitem{Djouadi} A.~Djouadi, J.~Kalinowski, and M.~Spira
  Comput. Phys. Commun. {\bf 108}, 56 (1998).
\bibitem{Kozlov:2005rj}  G.~A.~Kozlov, Phys.\ Rev.\ D {\bf 72}, 075015 (2005).
\bibitem{Ono:1983tf} S.~Ono, Acta Phys.\ Polon.\ B {\bf 15}, 201 (1984).
\bibitem{Cakir:2004nh}  O.~Cakir, R.~Ciftci, E.~Recepoglu and S.~Sultansoy,  Acta Phys.\ Polon.\ B {\bf 35}, 2103 (2004).
\bibitem{Hill} C.T.~Hill, E.H.~Simmons, Phys. Rept. {\bf 381}, 235 (2003) [Erratum-ibid. {\bf 390}, 553 (2004)].
\bibitem{us} D0 Collaboration, V.M. Abazov {\it et al.}, Phys. Lett. B {\bf 653}, 378 (2007).
\bibitem{Abazov:2006ez} D0 Collaboration, V.M. Abazov {\it et al.}, Phys. Lett. B {\bf 641}, 415 (2006); ``Erratum to
Search for Particles Decaying into a Z Boson and a Photon in ppbar Collisions at sqrt(s) = 1.96 TeV''.
\bibitem{run2det} D0 Collaboration, V.M. Abazov {\it et al.}, Nucl. Instrum. Methods Phys. Res. A {\bf 565}, 463 (2006).
\bibitem{lumi} T. Andeen {\it et al.}, FERMILAB-TM-2365 (2007).
\bibitem{Carena}
  M.S. Carena, A. Daleo, B.A. Dobrescu and T.M.P. Tait, Phys.\ Rev.\  D {\bf 70}, 093009 (2004).
\bibitem{pythia}
T. Sj{\"o}strand {\it et al.}, Computer Physics Commun. {\bf 135}, 238 (2001).
\bibitem{madevent}F. Maltoni and T. Stelzer, JHEP {\bf 0302}, 027 (2003).
\bibitem{EM_isolation} {\it Isolation} $= \frac{E_{\rm tot}(R<0.4) - E_{\rm EM}(R<0.2)}{E_{\rm EM}(R<0.2)}$, where $E_{\rm EM}~(R<0.2)$ is the EM energy within a cone of radius $R = \sqrt{(\Delta \phi)^{2} + (\Delta \eta )^{2}} =$ 0.2 and 
$E_{\rm tot}(R<0.4)$ is the total energy within a cone of radius R$=$0.4.
\bibitem{baur} U. Baur and E. Berger, Phys. Rev. D {\bf 47}, 4889 (1993).
\bibitem{baur_NLO} U. Baur, T. Han and J. Ohnemus, Phys. Rev. D {\bf 57}, 2823 (1998).

\bibitem{fisher} W. Fisher, FERMILAB-TM-2386-E (2007).
\bibitem{junk} T. Junk, Nucl. Instrum. Methods Phys. Res. A {\bf 434}, 435 (1999).
\end{thebibliography}
\end{document}